\documentclass[a4paper,11pt]{article}
\usepackage{graphicx}
\usepackage[english]{babel}
\usepackage[utf8]{inputenc}   
\usepackage{lmodern}
\usepackage[T1]{fontenc}     
\usepackage{geometry} 
\usepackage{graphicx} 
\usepackage{amsmath}
\usepackage{amsfonts}
\usepackage{amssymb}
\usepackage{multicol}
\usepackage{caption}
\usepackage{wrapfig}
\usepackage{array,multirow,makecell}
\usepackage{xcolor}
\usepackage{mathrsfs}
\usepackage[left,modulo]{lineno}
\usepackage[nottoc, notlof, notlof]{tocbibind}
\usepackage[toc,page]{appendix}
\usepackage{tikz}
\usepackage{feynmp}
\usepackage{subfigure}
\usepackage{rotating}
\usetikzlibrary{decorations.pathmorphing,decorations.markings}
\usetikzlibrary{arrows,shapes}
\usetikzlibrary{trees}
\usetikzlibrary{matrix,arrows} 	
\usepackage{tikz-qtree}			% For commutative diagram
\usetikzlibrary{positioning}				% For "above of=" commands
\usetikzlibrary{calc,through}				% For coordinates
\usetikzlibrary{decorations.pathreplacing}  % For curly braces
\usepackage{pgffor}							% For repeating patterns
\usetikzlibrary{decorations.pathmorphing}	% For Feynman Diagrams
\usetikzlibrary{decorations.markings}
\usepackage[squaren, Gray, cdot]{SIunits}

\definecolor{color1}{RGB}{ 85 107  47 }
\definecolor{bl}{RGB}{  65 105 225	}
\definecolor{rg}{RGB}{ 139   0   0 	}
\definecolor{vt}{RGB}{  46 139  87}
\usepackage{hyperref}
\hypersetup{ colorlinks = true, linkcolor =rg, urlcolor = bl, citecolor =vt }
\usepackage{fancyhdr}
\title{On the detection of neutrinos from solar flares using pion decay photons to provide a time window template.}
\usepackage{authblk}

\author[1]{G. de Wasseige}
\author[2]{K. Hanson}
\author[1]{N. van Eijndhoven}
\author[3]{P. Evenson}
\author[4]{K.-L. Klein}
\affil[1]{Interuniversity Institute for High-Energy, Vrije Universiteit Brussel, Brussels, Belgium\\
       gdewasse@vub.ac.be}
\affil[2]{WIPAC, University of Wisconsin - Madison, Madison, WI, USA}
\affil[3]{Department of Physics and Astronomy, University of Delaware, Newark, DE, USA}
\affil[4]{LESIA, Observatoire de Paris, Meudon, France}

\begin{document}
\date{}
\maketitle
\begin{center}\bf{Abstract}\end{center}
Since the end of the eighties and in response to a reported increase in the total neutrino flux in the Homestake experiment in coincidence with solar flares, solar neutrino detectors have searched for solar flare signals.
Even though these detectors have used different solar flare samples and analyses, none of them has been able to confirm the possible signal seen by Homestake.
Neutrinos from the decay of mesons, which are themselves produced in collisions of accelerated ions with the solar atmosphere would provide a novel window on the underlying physics of the hadronic acceleration and interaction processes during solar flares. Solar flare neutrino flux measurements would indeed help to constrain current parameters such as the composition of the accelerated flux, the proton/ion spectral index and the high energy cutoff  or the magnetic configuration in the interaction region.
We describe here a new way to search for these neutrinos by considering a specific solar flare sample and a data driven time window template which will improve the likelihood of neutrino detection.%}

\section{Evaluation of the solar flare signal in neutrino detectors}
\begin{figure}
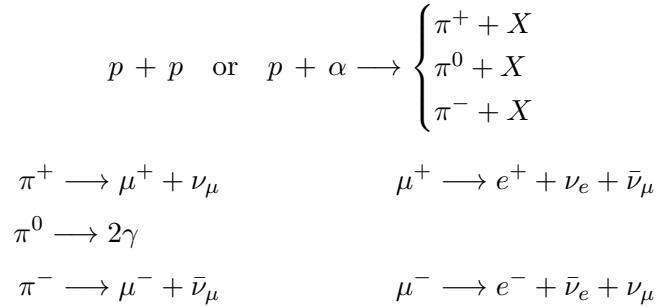

\[ p\, + \, p \quad \text{or} \quad p\, + \, \alpha \longrightarrow \begin{cases}
 \pi^+ + X &\\
  \pi^0 + X& \\
    \pi^- + X& \\
\end{cases}
\]

\[\pi^+ \longrightarrow \mu^+ + \nu_{\mu} \qquad\qquad\qquad \mu^+ \longrightarrow e^+ + \nu_e + \bar\nu_{\mu} \]
\[\pi^0 \longrightarrow 2 \gamma \qquad\qquad\,\,\,\,\,\quad\qquad\qquad\qquad \qquad\qquad\qquad \]
\[\pi^- \longrightarrow \mu^- + \bar{\nu}_{\mu} \qquad\qquad\qquad \mu^- \longrightarrow e^- + \bar{\nu}_e + \nu_{\mu} \]
%\[\pi^0 \longrightarrow 2 \gamma \qquad\qquad\,\,\,\,\,\quad\qquad\qquad\qquad \qquad\qquad\qquad \]
\caption{Sketch of the particle production processes \label{reaction}}
\[\]\end{figure}

Neutrinos from solar flares are produced through the interactions of accelerated protons/ions with the chromosphere and the subsequent decays of the produced charged pions as shown in Figure \ref{reaction}.

%%%\begin{center}
%$p_C = p\,(\text{from the chromosphere})\qquad  \,  \qquad\,p_{A} =  p\,(\text{accelerated})
 %\qquad(E_{p_{A}} \approx$ GeV$)$\end{center}
We have developed a Geant4 simulation of hadronic interactions in the active solar atmosphere in order to evaluate the flux of solar flare neutrinos we could expect at Earth. Using the proton spectra of R. J. Murphy and R. Ramaty \cite{modelA}, we find that the expected neutrino fluence covers an energy range from 10 MeV to more than 1 GeV as presented in Table \ref{table1}.
\begin{table}[h!]
\begin{center}
\begin{tabular}{c |  c | c}
  \hline
Energy range & Neutrino fluence at Earth &Spectrum \\
  \hline
10 - 100 MeV & 770 $\nu$ cm$^{-2}$& E$^0$ \\
100 - <1000 MeV & 783 $\nu$ cm$^{-2}$ & E$^{-2.3}$ \\
  \hline
\end{tabular}

\end{center}
\caption{\label{table1} Expected neutrino fluence at Earth and spectrum for one solar flare.}
\end{table}

We have studied the influence of the proton spectral index, the composition of the accelerated flux as well as the angular distribution of this flux. (Results will be presented at the \href{http://icrc2015.nl}{ICRC 2015}. )

\section{Solar flare selection}
\paragraph{} In addition to neutrinos, the interaction channels outlined above produce gamma-rays from the decay of neutral pions and bremsstrahlung of secondary electrons. The gamma-rays produced by neutral pion decay, which have an energy above 67 MeV, will be of great interest for neutrino searches because of their common production channels with respect to charged pions as indicated in Figure \ref{reaction}.

\paragraph{}Gamma-ray detectors such as RHESSI \cite{rhessi}, INTEGRAL \cite{integral}, CORONAS \cite{song} or the current FERMI-LAT \cite{lat} have been able to identify these interesting gamma-rays by searching for a pronounced pion-decay line in the spectrum of solar flares \cite{piondecay}. It is therefore possible to identify solar flares in which pion production occurs. We will call these specific solar flares "pion-flares".
These pion-flares are particularly interesting for neutrino searches since they are a guaranteed source of neutrinos. Using only these pion-flares as a sample for their analyses, neutrino detectors from this generation such as IceCube \cite{icecube}, Antares \cite{antares} and SuperKamiokande \cite{superkamiokande} or the next generation detectors Pingu \cite{pingu}, KM3NeT \cite{km3net}, Orca \cite{orca} and Juno \cite{juno} could optimise their sensitivity for the detection of related neutrino activity.
  
\section{Time window template}
We describe here a time window template for neutrino searches which could be used in the case of observations of pion-decay gamma-rays during the impulsive phase of the flare. However, also long duration events are interesting candidates for neutrino searches, as indicated below.  Apart from providing a background reduction, these time windows will also allow a stacked analysis similar to GRB neutrino detection via time profile stacking \cite{nick_grb}. 
\subsection{Impulsive phase of the pion-flares}
\begin{figure}[h!]
    \centering
    \includegraphics[width=0.5\textwidth]{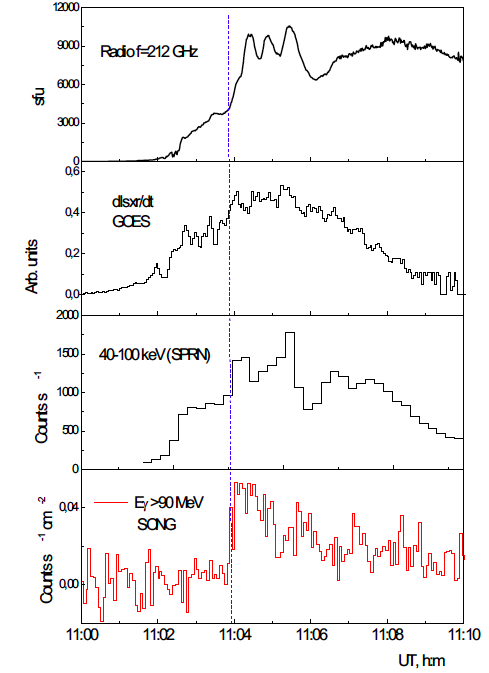}
    \caption{Time series of the impulsive phase of a pion-flare for the 28 October 2003 event \cite{icrc2011} \label{gamma_idea}.}
\end{figure}

V. Kurt et al. have analyzed five events tagged as pion-flares \cite{icrc2011}. They have shown that in all of these events, the FWHM of the pion-decay burst in the impulsive phase of the flare does not exceed 4 minutes even though the duration of the rise time up to the maximum and the total duration of the pion-decay gamma-ray burst vary from flare to flare. An example  of this is presented in Figure \ref{gamma_idea}.
Because of the common production channels of neutral and charged pions, the short time window during the impulsive phase of the flare in which gamma-rays from the decay of neutral pions are observed constitutes a useful template for the time window in which neutrinos will be produced. It therefore means that using the starting time of this gamma-ray emission and opening a window of a maximum of 4 minutes will allow a search for solar flare neutrinos with a significantly reduced background compared to previous searches (e.g. \cite{bahcall-flares,sl,sno2}). 

Considering the facts outlined above, both flare-per-flare and flare stacking searches will benefit from reduced background conditions.

\subsection{Long duration events}
In some events gamma-ray emission by pion-decay photons was found to persist during several hours, much longer than the impulsive phase of the flare \cite{Kan,fermi_lde}. Long duration events are therefore additional interesting candidates for neutrino searches. However, large scale neutrino detectors, like e.g. IceCube, should restrict their time window to a maximum of about 20 minutes in order to maintain a reasonable signal to noise ratio.

\section{Preliminary list of proposed solar flares}
We present here two initial lists of solar flares which could be used for neutrino searches. These lists are directly inspired from \cite{pionflarefermi}. 
\subsection{Pion-decay gamma-ray observations in the impulsive phase}
Gamma-rays from pion-decay have been observed in the impulsive phase of all of the listed flares. Start and duration information have been obtained from Fermi LAT.

\begin{table}[h!]
\begin{center}
\begin{tabular}{c | c | c }
  \hline
 Date & Start  (in UT, hh:mm)& Duration (minutes) \\
  \hline
  2010 Jun 12&00:55& 0.8\\
  2011 Aug 9& 08:01 & 3.3 \\
  2011 Sep 6& 22:17 & 0.2 \\
  2011 Sep 24&09:34& 0.8\\
  2012 Jun 3& 17:52 & 0.6 \\
  \hline
\end{tabular}
\caption{First proposed solar flare list for impulsive phase neutrino searches. }
\end{center}
\end{table}

\subsection{Extended pion-decay gamma-ray emissions}
As previously mentioned, the following solar flares emitted gamma-rays from pion-decay for a time longer than the impulsive phase duration. 
\begin{table}[!h!]
\begin{center}
\begin{tabular}{c | c | c || r | c | c }
  \hline
 Date & Start & Duration  & Date & Start  & Duration \\
   &  (in UT, hh:mm)& (minutes) & & (in UT, hh:mm)& (minutes)\\
  \hline
2011 Mar 7&20:15& 25    &2012 Mar 5& 04:12 & 49            \\  
                   &23:26& 36	&	 & 05:26 & 71	\\ \cline{1-3}
 2011 Mar 8&02:38& 35		&& 07:23 & 28	\\ \cline{4-6}
                  &05:49& 35	&	2012 Mar 7& 00:46 & 31	\\ \cline{1-3}
  2011 Jun 2& 09:43 & 45	&	& 00:46 &60	 \\ \cline{1-3}
  2011 Jun 7& 07:34 & 53	& & 03:56 & 32		 \\\cline{1-3}
    2011 Aug 4&04:59& 34	&& 07:07 & 32		\\\cline{1-3}
  2011 Sep 6& 22:13 & 35 		&  & 10:18 & 32	\\\cline{1-3}
  2011 Sep 7& 23:36 & 63 	&	  & 13:29 & 32	\\\cline{1-3}
  2012 Jan 23& 04:07 & 51 	&  & 19:51 & 25 		\\\cline{4-6}
 		    & 05:25 & 69 	&	  2012 Mar 9& 06:52 & 34	\\
		     & 07:26 & 16 	&& 08:28 & 34 		\\\cline{4-6}
		     & 08:47 & 35 	&	  2012 May 17& 02:18 & 22	\\\hline
  2012 Jan 27& 21:13 & 24	&	   2012 Jun 3& 17:40 & 23	 \\ \cline{4-6}
   &&& 2012 Jul 6& 23:19 & 52 \\

  \hline
\end{tabular}
\caption{First proposed solar flare list for extended neutrino searches. }
\end{center}
\end{table}
\section{Conclusion and Outlook}

The current number of identified pion-flares constitutes a workable starting sample for neutrino searches, even though none of the actual high-energy gamma-ray telescopes are continuously observing the Sun and consequently some identifications of solar flares as pion-flares may be missed.
Furthermore, the short time window of the pion-decay gamma-ray burst in the impulsive phase of the flares enables a drastic reduction of the time integrated background in neutrino detectors and also allows a stacked analysis. 
Using this flare sample and searching for a neutrino signal in the short time windows as outlined here, we expect current neutrino detectors to be able to detect a solar flare signal in an energy range from several MeV to GeV.

\bibliographystyle{plain}	
\bibliography{biblio}

\begin{thebibliography}{10}

\bibitem{integral}
K.~Hurley A.~Rau, A. V.~Kienlin and G.~G. Licht.
\newblock The 1st {I}{N}{T}{E}{G}{R}{A}{L} {S}{P}{I}-{A}{C}{S} gamma-ray burst
  catalogue.
\newblock {\em A\&A}, 438(3), 2005.

\bibitem{bahcall-flares}
J.~N. Bahcall.
\newblock Solar {F}lares and {N}eutrino {D}etectors.
\newblock {\em Phys. Rev. Lett.}, 61(23):2650--2652, 1988.

\bibitem{piondecay}
E.~L. Chupp and J.~M. Ryan.
\newblock High energy neutron and pion-decay gamma-ray emissions from solar
  flares.
\newblock {\em A\&A}, 9(1):11--40, 2009.

\bibitem{antares}
Antares Collaboration.
\newblock http://antares.in2p3.fr.

\bibitem{icecube}
IceCube Collaboration.
\newblock https://icecube.wisc.edu.

\bibitem{juno}
Juno Collaboration.
\newblock http://english.ihep.cas.cn/rs/fs/juno0815/.

\bibitem{km3net}
K{M}3{N}e{T} Collaboration.
\newblock http://www.km3net.org/home.php.

\bibitem{superkamiokande}
Super{K}amiokande Collaboration.
\newblock http://www-sk.icrr.u-tokyo.ac.jp/sk/index-e.html.

\bibitem{sno2}
B.~Aharmim et~al.
\newblock A search for astrophysical burst signals at the {S}udbury {N}eutrino
  {O}bservatory.
\newblock {\em Astropart.Phys.}, 55:1--7, 2014.

\bibitem{pionflarefermi}
M.~Ackermann et~al.
\newblock High-energy gamma-ray emission from solar flares: summary of
  {F}{E}{R}{M}{I}-{L}{A}{T} detections and analysis of two m-class flares.
\newblock {\em Ap{J}}, 787(15), 2014.

\bibitem{fermi_lde}
M.~Ajello et~al.
\newblock Impulsive and long duration high-energy gamma-ray emission from the
  very bright 2012 march 7 solar flares.
\newblock {\em Ap{J}}, 789:20--35, 2014.

\bibitem{pingu}
M.G.~Aartsen et~al.
\newblock Letter of {I}ntent: The {P}recision {I}ce{C}ube {N}ext {G}eneration
  {U}pgrade ({P}{I}{N}{G}{U}).
\newblock {\em arXiv:1401.2046 [physics.ins-det]}, 2014.

\bibitem{rhessi}
R.~P.~Lin et~al.
\newblock The {R}euven {R}amaty {H}igh-{E}nergy {S}olar {S}pectroscopic
  {I}mager ({R}{H}{E}{S}{S}{I}).
\newblock {\em Solar {P}hys.}, 210(1):3--32, 2002.

\bibitem{orca}
U.F.~Katz et~al.
\newblock The {O}{R}{C}{A} option for {K}{M}3{N}e{T}.
\newblock In {\em XV Workshop on Neutrino Telescopes}, 2013.

\bibitem{lat}
W.~B.~Atwood et~al.
\newblock The {L}arge {A}rea {T}elescope on the {F}ermi {G}amma-ray {S}pace
  {T}elescope {M}ission.
\newblock {\em Ap{J}}, 697(1071), 2009.

\bibitem{sl}
K.~S. Hirata, T.~Kajita, T.~Kifune, K.~Kihara, M.~Nakahata, K.~Nakamura,
  S.~Ohara, Y.~Oyama, N.~Sato, M.~Takita, Y.~Totsuka, and Y.~Yaginuma.
\newblock Search for correlation of neutrino events with solar flares in
  {K}amiokande.
\newblock {\em Phys. Rev. Lett.}, 61(23):2653--2656, 1988.

\bibitem{Kan}
G.~Kanbach, D.L. Bertsch, C.E. Fichtel, R.C. Hartman, S.D. Hunter, D.A.
  Kniffen, P.W. Kwok, Y.C. Lin, J.R. Mattox, and H.A. Mayer-Hasselwander.
\newblock Detection of a long-duration solar gamma-ray flare on june 11, 1991
  with {E}{G}{R}{E}{T} on {C}{O}{M}{P}{T}{O}{N}-{G}{R}{O}.
\newblock {\em A\&A}, 97:349--353, 1993.

\bibitem{icrc2011}
V.~Kurt, B.~Yushkov, and V.~Grechnev.
\newblock The onset time of the pion-decay gamma-ray emission of major solar
  flares.
\newblock In {\em 32nd International Cosmic Ray Conference}, 2011.

\bibitem{song}
S.~N. Kuznetsov, V.~G. Kurt, B.~Yu. Yushkovand, I.~N. Myagkova, V.~I. Galkin,
  and K.~Kudela.
\newblock {\em Protons Acceleration in Solar Flares: The Results of the
  Analysis of Gamma-emission and Neutrons Recorded by the {S}{O}{N}{G}
  Instrument Onboard the {C}{O}{R}{O}{N}{A}{S}-{F} Satelliteration in Solar
  Flares: The Results of the Analysis of Gamma-emission and Neutrons Recorded
  by the {S}{O}{N}{G} Instrument Onboard the {C}{O}{R}{O}{N}{A}{S}-{F}
  Satellite}, volume 400.
\newblock Astrophysics and Space Science Library, 2014.

\bibitem{modelA}
R.~J. Murphy and R.~Ramaty.
\newblock Solar-flare neutrons and gamma-rays.
\newblock {\em Adv. {S}pace {R}ES.}, 4(7), 1984.

\bibitem{nick_grb}
N.~van Eijndhoven.
\newblock On the observability of high-energy neutrinos from gamma ray bursts.
\newblock {\em Astropart.Phys.}, 28(540), 2008.

\end{thebibliography}
\end{document}